\documentclass[epj]{svjour}
\usepackage{amssymb, amsfonts, graphicx, hyperref, psfrag, epsfig}
\usepackage{cite}
\usepackage{subfigure}
\newcommand{\be}{\begin{equation}}
\newcommand{\ee}{\end{equation}}
\newcommand{\bea}{\begin{eqnarray}}
\newcommand{\eea}{\end{eqnarray}}
\newcommand{\ba}{\begin{array}}
\newcommand{\ea}{\end{array}}

\def\a{\alpha}

\def\r{\rho}

\def\O{\Omega}

\csname @addtoreset\endcsname{equation}{section}

\begin{document}
\title{Symmetry Energy from Holographic QCD}
\author{Yunseok Seo and  Sang-Jin Sin\inst{1}
\thanks{\emph{Present address:} Insert the address here if needed}%
}                     
\offprints{}          
\institute{Department of Physics, Hanyang University,
Seoul 133-791, Korea}
\date{Received: 2013.9.15 / Revised version: date}
%
\abstract{We review the symmetry energy in the context of AdS/CFT correspondence. After constructing $D$ brane configurations   corresponding to dense system in boundary theory, we calculate the symmetry energy by solving DBI action of $D$ branes in confining and deconfining phase. We conclude the density dependence of the symmetry energy  has scaling law whose power  depends only on dimensionality of the branes and space-time. 
} 
\maketitle
%
 
%
%
%

Nuclear symmetry energy is one of key words in nuclear physics as well as  in astrophysics. Its density dependence is essential to understand neutron star properties. It is surprising that such an important quantity is still 
 poorly understood after 80 years of its definition. See references \cite{sE1,sE2,sE3,sE5,sE6,sE7,Lee:2010sw,XLCYZ,Che05a} for its reviews  and  also see \cite{sE3} for experimental side.
  Such delay is due to  the lack of reliable calculational tool for   strongly interacting system especially in the presence of the chemical potential.
  Therefore  the most urgent task in nuclear physics is to invent a calculational tool    for  such system. 
  
  Recently gauge gravity duality ~\cite{Maldacena:1997re, Gubser:1998bc, Witten:1998qj}  is proposed as a new tool to calculate 
  strongly interacting. Strongly interacting quantum mechanical system is replaced by the classical gravity, making the problem much easier. 
The fist application of this duality to QCD  came by Son and his collaborators 
\cite{Son} for the heavy ion collision. Hydrodynamics flow analysis need to assume 
unusually small viscosity to fit the experimental data, while the perturbative 
result gives very large viscosity $\sim 1/g^4\log g$ in its validity regime $g<<1$. 
In \cite{Son}, the authors noticed that if gravity dual of thermalized quark gluon plasma  exists as a black hole background, both  viscosity which is proportional to the absorption cross section  and the system entropy   
should be proportional to black hole area, therefore  their ratio is a constant which turns out to be $1/4\pi$. The smallness of this number $\sim 0.1$  and its (almost) universality  is the key of the success story.  

For the confining phase, the  story is more subtle. 
The meson spectrum in holographic dual does not follow Regge trajectory unless 
one introduces long string mimicking the old QCD flux string. 
However, low lying meson spectrum could be fit very well (with 10$\%$ error!) with simplest models with only two parameters \cite{hQCD}.  The density and temperature dependence of the meson spectrum does not follow the  naive expectation: in many of the  holographic models, the meson spectrum increases as   temperature or density increases \cite{KSZ, Erdmenger}. 
Such expectation \cite{Brown} is coming from the smooth  interpolation of the 
chiral condensation from zero to critical density where it vanishes. 
At this moment, it is fare to say that 
we need to open our mind without relying on too much `intuition': mother nature did not revealed much about  her secret  hard core to us upto now. 
  
  The purpose of this article is to review  the result of its application to the nuclear symmetry energy based on a few models mimicking QCD~\cite{D4D6_03, hQCD} and following the way to treat the  dense matter in confined phase suggested in \cite{Seo:2008qc, KSS2010}.
We will find that the symmetry energy is increasing with the total charge $Q$, showing that the symmetry energy of our system has a stiff dependence  on the density. 
   
  Before we go further, it is better to specify what is pros and cons of this method.  First limitation comes from its large $N$ nature. 
  While many result in large N limit turns out to be valid for finite value of N, 
if the leading term of a physical quantity comes only from 1/N order or higher, its behavior can be very different from the  real QCD.  
  Secondly it is based on super symmetric theory and therefore one may worry  the contribution from unnecessary matter components.  Thirdly, it is higher dimensional theory and therefore it contains a tower of  Kaluza-Klein spectrum which do not decouple. In principle, we need to restrict ourselves to low energy spectrum. However, there is a study claiming that one gets better result if one consider all such infinite tower in the meson spectrum \cite{Hong:2007ay}.
  
 Therefore the best policy at this moment is that 
  while we should try to utilize it, we should also 
  look for a physical quantity  that is universal,  like $\eta/s$ \cite{Policastro:2001yc}. 
  In this spirit, we will show that 
  the result is rather insensitive to the shape of the brane embedding and metric deformation.  On the other hand, we will see that the scaling exponent depends on the dimensionality of the 
color and flavor branes. We call such discrete dependence of the scaling dimension as the universality class 
of the symmetry energy. 

\section{Symmetry energy in holographic QCD}
The symmetry energy of nuclear matter is defined as the energy per nucleon required
to change isospin symmetric nuclear matter to pure neutron matter.
The Bethe-Weizs\"{a}cker  formula represents the amount of binding energy that a nucleus has to lose when the numbers
of protons and neutrons are different.
The semi-empirical mass formula based on the liquid drop model has the form:
\bea
&&E_{\rm B}=a_{\rm v}\, A-a_a\, (N-Z)^2/A -a_c\,Z^2/A^{1/3} \nonumber \\
&&~~~~~~~~-a_s\, A^{2/3}\pm a_\delta/A^{3/4}\, .\label{BWf}
\eea
Here $Z$ ($N$) is the number of protons (neutrons) in a nucleus.
The first term is called the volume energy where $A$ is the total nucleon number.
The second  term defines the symmetry energy of nuclear matter.
If there were no Coulomb repulsions between protons, we would expect to have equal number of neutrons and protons in
nuclei in general.
The term with $a_c$ accounts for the Coulomb interaction between protons in the nucleus.
The last two terms represent the surface energy and pairing effect, respectively.

Due to the isospin invariance, iso-scalar quantities in a nuclear system
are function of only even powers of the asymmetry factor $\alpha$ defined by
 $\alpha\equiv (N-Z)/A$. Then we can express the energy density per nucleon $E(\rho,\alpha)$ as

\be\label{E0}
E(\rho,\a)=E(\rho,0) + E_{\rm sym}(\rho)\,\a^2 +O(\a^4).
\ee
The nuclear symmetry energy is defined as the coefficient $E_{\rm sym}(\rho)$ in (\ref{E0}). 
There is no term which is odd power in $\a$ due to the exchange symmetry between protons and neutron in nuclear matter.  It is a energy cost per nucleon to deviate the line $Z=N$.

The purpose of this work is to calculate the symmetry energy (\ref{E0}) from the AdS/CFT correspondence. To to this, we first  construct D brane configuration which corresponds to the system with finite density. After construct D-brane configuration, we calculate free energy of the system which is related to the energy density at the boundary system (\ref{E0}). Once we get total energy density of the system, we can easily calculate the symmetry energy as we will describe later.

We first consider $D4$ brane background and use $D6$ branes as probes. In this background we calculate nuclear symmetry energy in confining and deconfining phase. Then we generalize the result to other probe brane and background. The original idea of AdS/CFT correspondence is based on $D3$ brane and it's near horizon geometry. But it can be generalized to the other $D$ branes and we will apply the concept of AdS/CFT to the system of $D4$ brane with proper compactification.

\begin{figure}[ht!]
\begin{center}
\includegraphics[height=0.25 \textwidth]{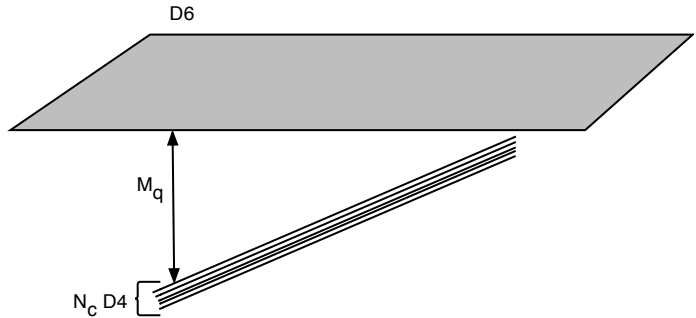}
\caption{$D4/D6$ brane configuration in weak string coupling limit.}\label{fig:D4D60}
\end{center}
\end{figure}

The key idea of AdS/CFT is that the gauge theory with strong interaction can be related to the classical gravity in higher dimension. The gauge theory is living at the boundary of higher dimensional bulk. From the 10 dimensional string theory, the massless excitation of fundamental strings on $N_C$ $D4$ branes can be identified by gauge field with $SU(N_C)$ gauge group which is color symmetry group at the boundary. The flavor symmetry can be introduced by putting another (probe) brane. If the boundary theory is in $3+1$ dimension, proper probe brane is $D6$. This probe brane carries flavor degrees of freedom. The fundamental string which connect $N_C$ $D4$ brane and probe $D6$ brane carries color and flavor index. It can be identified to quarks in boundary theory. The length of fundamental strings corresponds to the mass of quarks, hence the distance between $D4$ brane and $D6$ brane can be identified by the mass of quark, $M_q$. The $D$ brane configuration which we concerned is drawn in Figure \ref{fig:D4D60}.

\begin{figure}[!ht]
\begin{center}
\subfigure[]{\includegraphics[angle=0, width=0.45\textwidth]{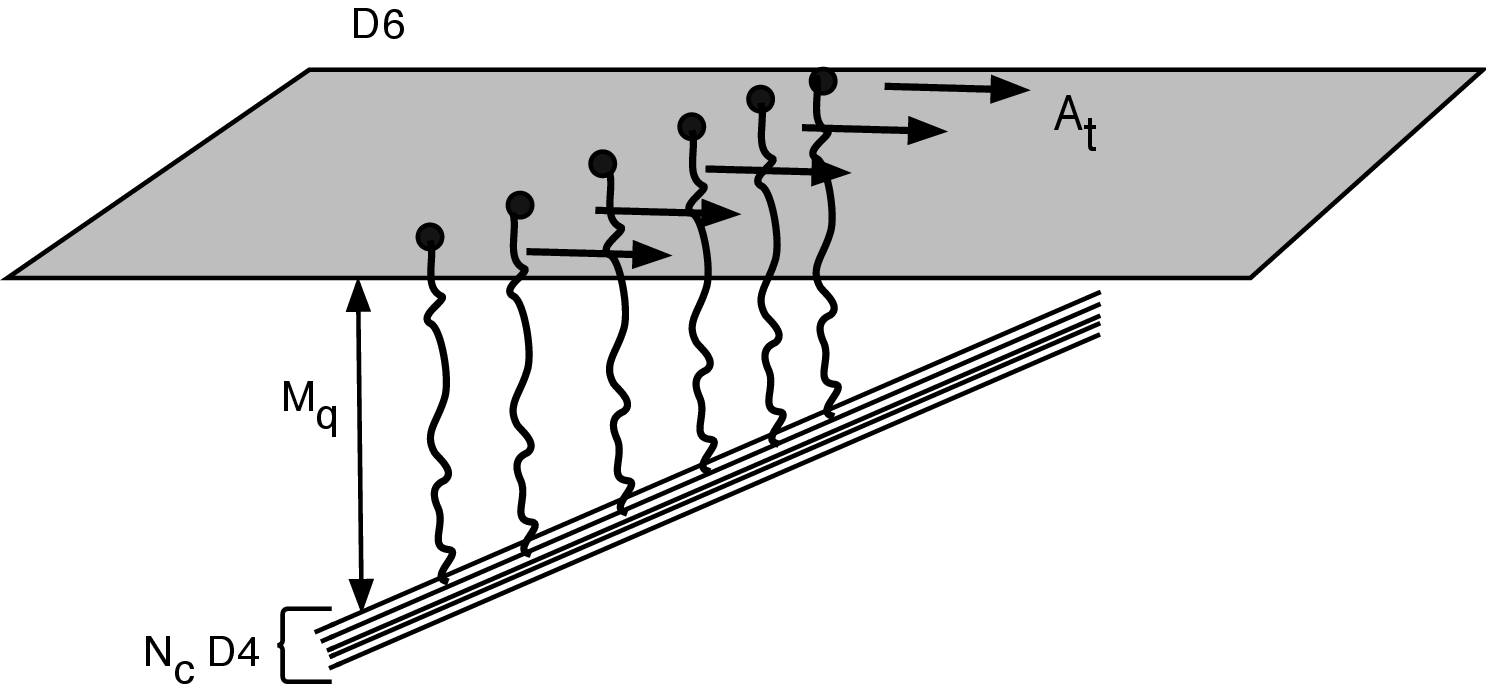}}
\subfigure[]{\includegraphics[angle=0, width=0.45\textwidth]{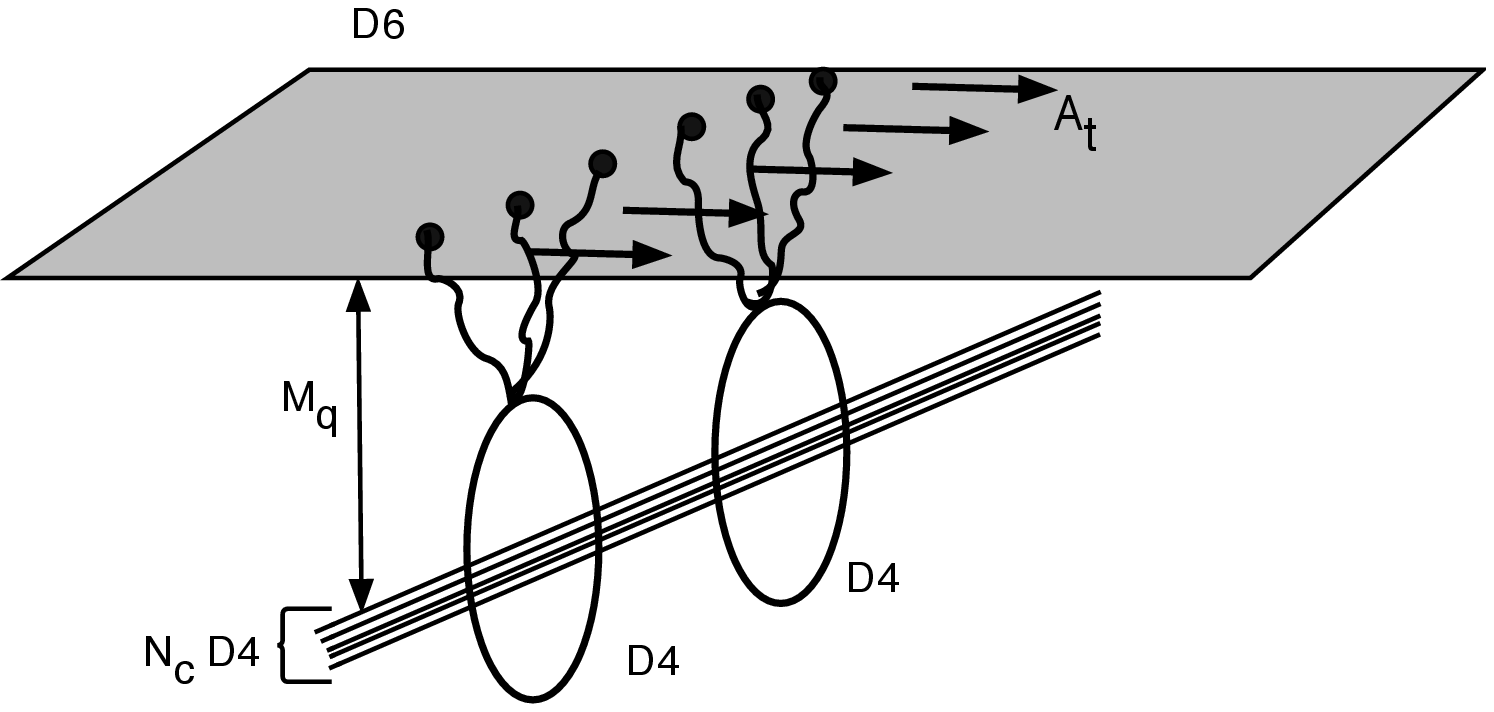}}
\caption{Schematic figure of D-brane configuration for finite density system. \label{fig:D4D601}}
\end{center}
\end{figure}

From the AdS/CFT correspondence, the boundary global symmetry is related to the local symmetry in the bulk. For example, the $U(1)$ global current at the boundary is related to the $U(1)$ gauge field in the bulk;
\be
A_{\mu} \leftrightarrow J^{\mu} = <\bar{\psi} \gamma^{\mu}\psi>.
\ee
Therefore, if we turn on $A_t$ field on the probe brane, the boundary current $J^{0} =<\psi^{\dagger}\psi> =<Q>$ is nothing but the expectation value of density operator at the boundary. But this electric field on the probe brane needs point like sources which couple to the field. The stringly object which corresponds to the point like object is an endpoint of fundamental string. There are two way to put fundamental strings on probe $D6$ brane. One is putting fundamental strings such that they connect $N_C$ $D4$ branes and probe $D6$ brane. The other way is introducing spherical $D4$ brane and connecting fundamental strings from the spherical $D4$ brane to probe $D6$ brane. The schematic figures are drawn in Figure \ref{fig:D4D601}.

In Figure \ref{fig:D4D601} (a), the fundamental strings create density. As we discussed before, the interpretation of fundamental string at the boundary theory is quark. Therefore, the system for Figure \ref{fig:D4D601} (a) can be understood as a quark matter system. On the other hand, the object which generates density in Figure \ref{fig:D4D601} (b) is $D4$ brane with fundamental quarks. The preserve flux from $N_C$ $D4$ brane, the number of fundamental quark on each spherical $D4$ brane should be $N_C$  \cite{Witten:1998xy}. Therefore, the object which generates density is $D4$ brane with $N_C$ fundamental quarks so called `baryon vertex'. From the boundary theory point of view, this object looks baryon because it is nothing but certain bound state of $N_C$ quark and the boundary system for Figure \ref{fig:D4D601} (b) is a nuclear matter system.

Now let's take large $N_C$ limit. In this limit, the number of color $D4$ brane becomes large. Due to the tension of $D4$ brane, large number of color brane can affect background geometry. Here we keep number of $D6$ brane is one of two such that it does not back react to the background geometry. One solution of large number of $D4$ brane is near horizon limit of black $D4$ branes;
\bea\label{deconfbg}
ds^{2}
&=&\left(\frac{U }{R }\right)^{3/2}\left(-f(U) dt^{2} +d\vec{x}^2 +dx_4^2  \right) \cr
 & &+\left(\frac{R}{U }\right)^{3/2}\left( \frac{dU^2}{ f(U)} +U^2 d\Omega_4^2\right) \cr
e^\phi&=&g_s\left(\frac{U }{R }\right)^{3/4},\quad F_4 =\frac{2\pi N_c}{\Omega_4}\epsilon_4, \;\;\cr
  f(U)&=& 1-\Big(\frac{U_{0}}{U}\Big)^{3}, \;\; R^3=\pi g_s N_C l_s^3.
\eea
There is a horizon at $U=U_0$, and the Hawking temperature and the horizon radius are related by \cite{Kruczenski:2003uq}
\be
U_0 =\frac{16 \pi^2}{9} R^3 T^2.
\ee
We identify this Hawking temperature to the temperature of boundary theory.

We can obtain another background solution by taking double Wick rotation by $t \leftrightarrow i x_4$ and $ x_4 \leftrightarrow i\tau$. Because Einstein tensor and curvature scalar does not change under Wick rotation, one can easily expect that Wick rotated geometry is also a solution of Einstein equation. The metric of double Wick rotated geometry is
\begin{eqnarray}\label{confbg}
ds_{D4}^{2}
&=&\left(\frac{U }{R }\right)^{3/2}\left(\eta_{\mu\nu}dx^\mu dx^\nu + f(U) dx_4^{2} \right)\cr 
& & +\left(\frac{R}{U }\right)^{3/2}\left( \frac{dU^2}{ f(U)} +U^2 d\Omega_4^2\right) \cr
e^\phi&=&g_s\left(\frac{U }{R }\right)^{3/4},\quad F_4 =\frac{2\pi N_c}{\Omega_4}\epsilon_4, \;\;\cr 
 f(U)&=&
1-\Big(\frac{U_{KK}}{U}\Big)^{3}, \;\; R^3=\pi g_s N_c l_s^3. 
\label{adsm}
\end{eqnarray}
This geometry does not have black hole horizon  and we believe that the temperature of boundary theory is zero. The geometry ends up at $U =U_{KK}$. It provides scale of the background theory which is Kaluza-Klein mass scale $M_{KK}$ which is defined as inverse radius of the $x_4$ direction:
$M_{KK}=\frac{3}{2}\frac{U^{1/2}_{KK}}{R^{3/2}}$.
The bulk parameters, $U_{KK}, g_s, R$ and the   gauge theory parameters
$M_{KK}, g^2_{YM}, \lambda:=g_{YM}^{2}N_{C}$ are related by
\be\label{consts1}
g_s=\frac{\lambda}{2\pi l_sN_c M_{KK}}, \quad U_{KK}=\frac{2}{9}\lambda M_{KK} l_s^2,
\quad R^3=\frac{\lambda l_s^2}{2M_{KK}} .
\ee

\begin{figure}[!ht]
\begin{center}
\subfigure[]{\includegraphics[angle=0, width=0.45\textwidth]{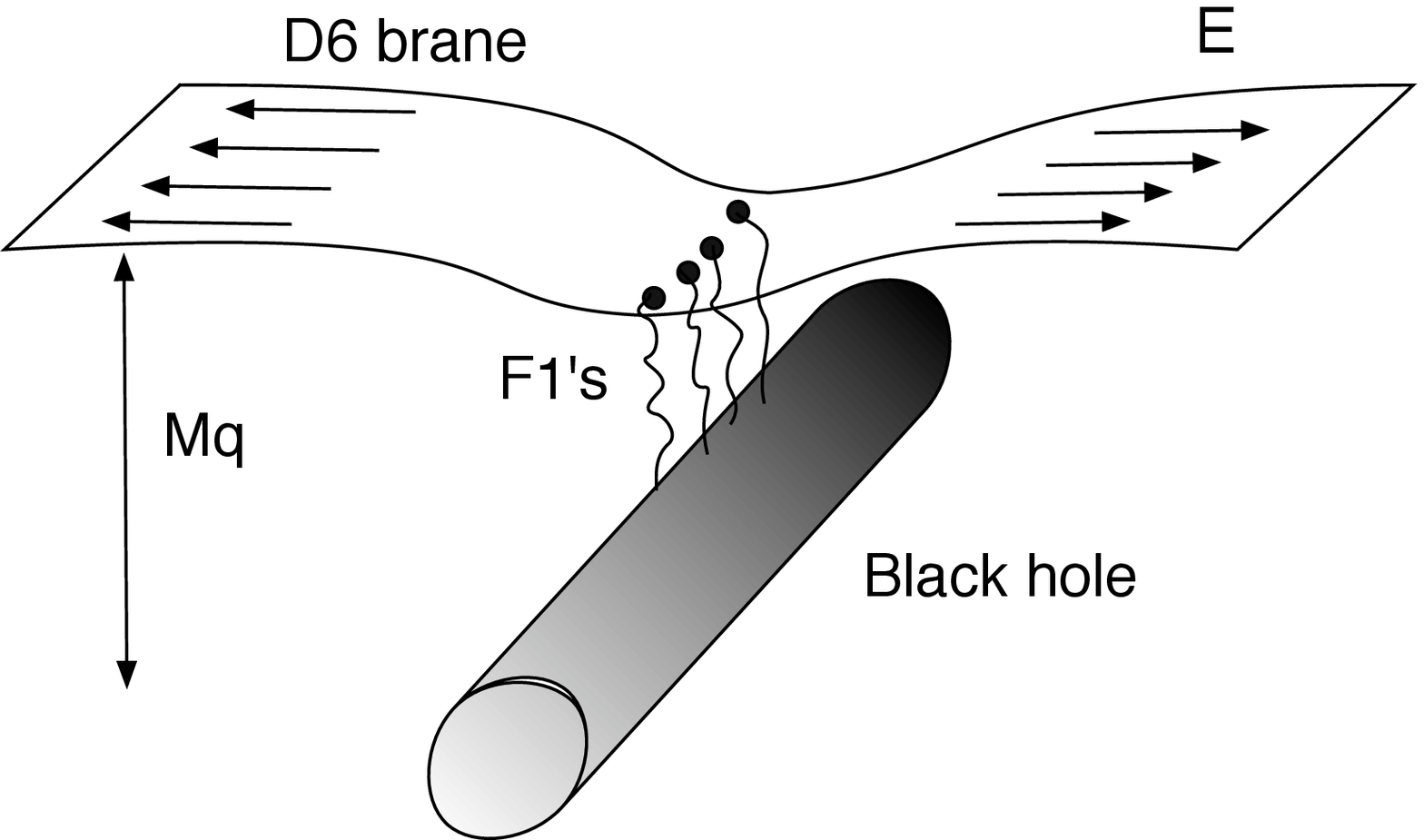}}
\subfigure[]{\includegraphics[angle=0, width=0.45\textwidth]{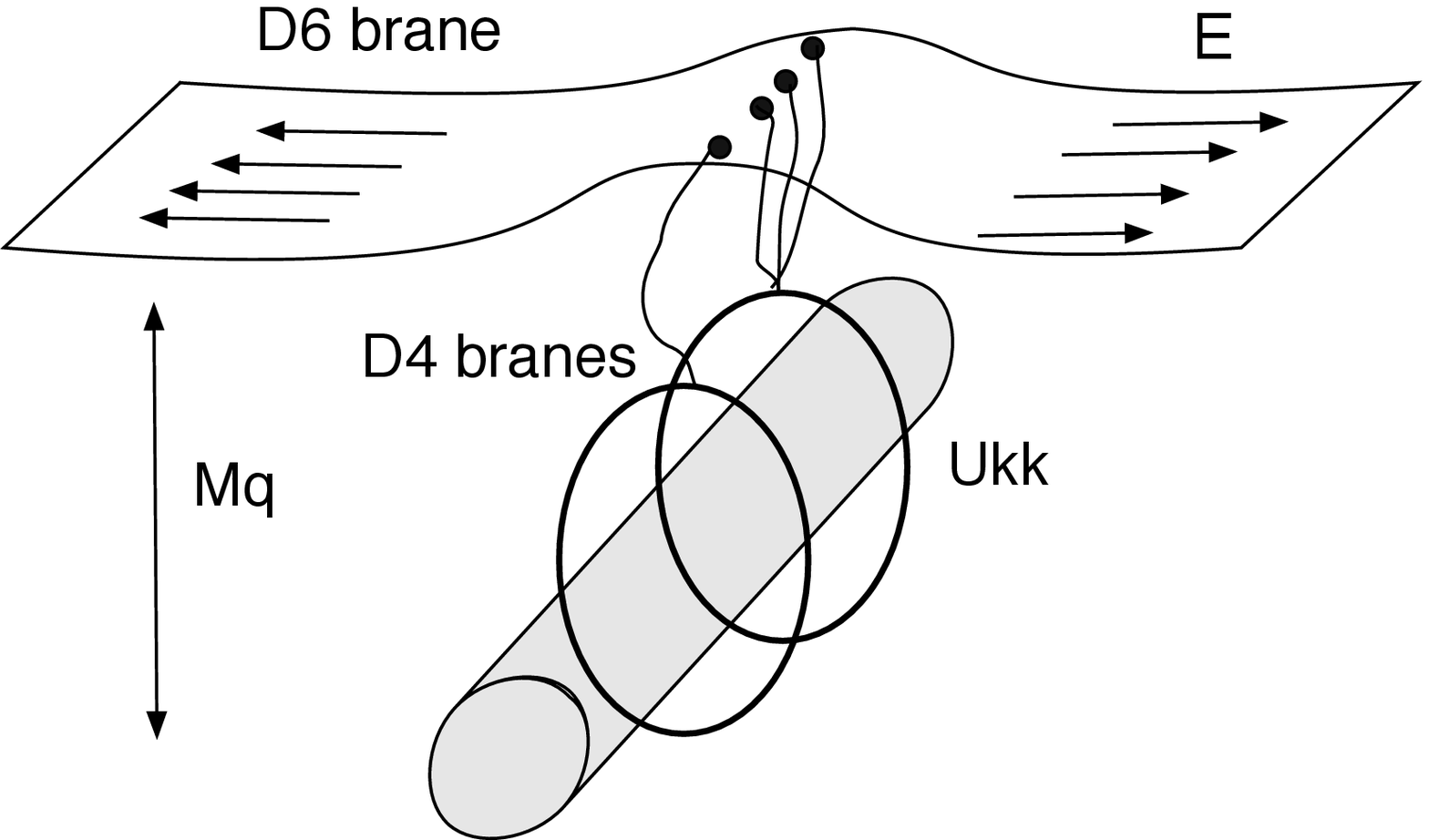}}
\caption{Schematic figure of D-brane configuration for finite density system. (a) for quark matter system (b) for nuclear matter system.\label{fig:D4D602}}
\end{center}
\end{figure}

Now we return to the $D$ brane configurations. In finite temperature geometry (\ref{deconfbg}) does not allow baryon vertex \cite{Seo:2008qc}. Therefore, the only way to introduce finite density to the system is connect fundamental strings between probe brane and black hole horizon. In this case, boundary system can be identified by the system with freely moving quarks under finite temperature, i. e. `quark matter system'.  On the other hand double Wick rotated geometry(or confined geometry) does not allow fundamental quark to generate density because there is no object where the other endpoint of fundamental strings attached on. therefore, baryon vertex should be introduce to generate density. In this case, the boundary system is that of baryons i. e. `nuclear matter system'.  The schematic figures are drawn in Figure \ref{fig:D4D602}.

The $D$ brane configurations in Figure \ref{fig:D4D602} are not stable due to the difference of tension of the different $D$ branes and fundamental strings. The dynamics of $D$ branes is governed by Dirac-Born-Infeld(DBI) action;
\be\label{Sdbi}
S_{Dq} =\mu_q \int d^{q+1} x \,\, \sqrt{\det(g_{MN} +2 \pi \alpha' F_{MN})},
\ee
where $\mu_q$ is tension of $Dq$ brane, $q$ is spatial dimension of probe brane and $g_{MN}$ is induced metric which defined on the probe brane. 

By solving the equation of motion of DBI action, we can get embedding solution of the probe brane. To solve the equation of motion we have to impose proper boundary condition. In the case of quark matter system(Figure \ref{fig:D4D602} (a)), tension of fundamental strings are always bigger than that of probe $D6$ brane. It means that the length of fundamental strings shrinks to zero hence probe $D6$ brane fall into the black hole horizon.  Then, we need to impose boundary condition for probe $D6$ brane at the horizon. This boundary condition is determined by the regularity condition. The equation of motion is written in terms of induced metric of the embedding. Near black hole horizon, time component of the induced metric vanishes and equation of motion seems to be diverge. To prevent this divergence we have to impose the boundary condition such that the probe brane touch the black hole horizon perpendicularly.

\begin{figure}[!ht]
\begin{center}
\subfigure[]{\includegraphics[angle=0, width=0.4\textwidth]{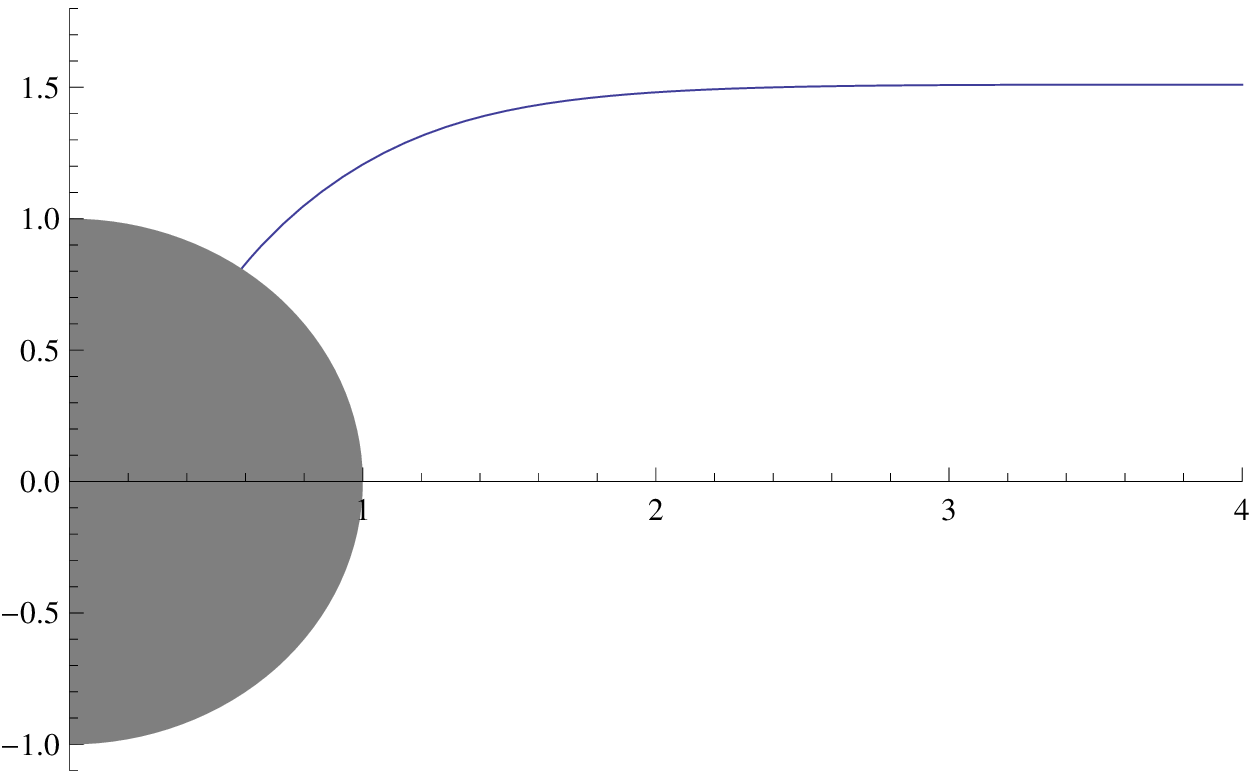}}
\subfigure[]{\includegraphics[angle=0, width=0.4\textwidth]{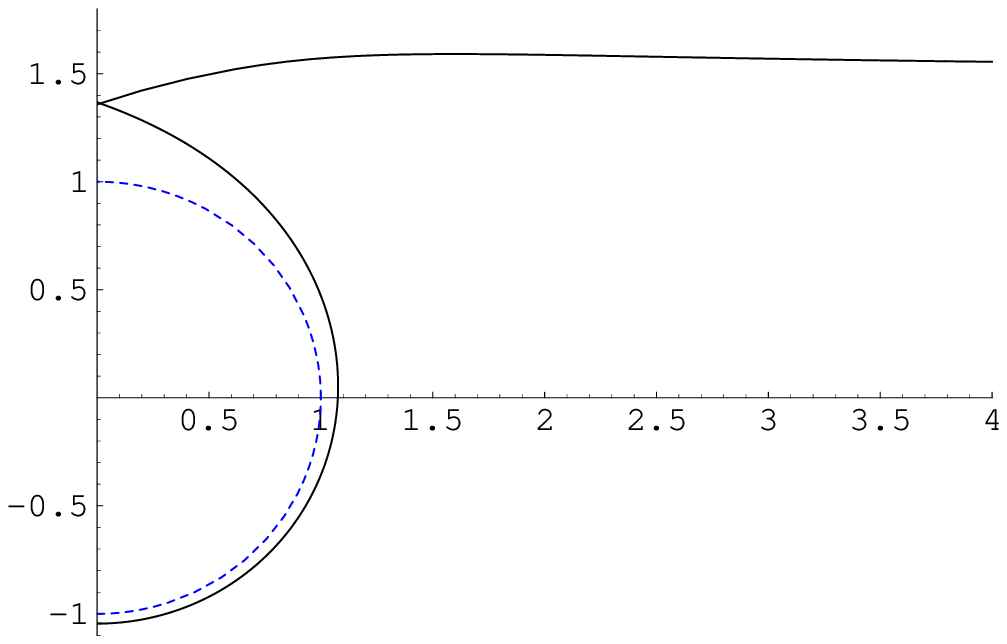}}
\caption{Solution of probe brane  (a) for quark matter system (b) for nuclear matter system.\label{fig:D4D603}}
\end{center}
\end{figure}

On the other hand, the $D$ brane configuration for nuclear matter system(Figure \ref{fig:D4D602} (b)) does not have divergence in the equation of motion. Instead of requiring regularity condition, we have to impose force balance condition. The tension of fundamental strings is alway bigger than that of probe $D6$ and spherical $D4$ brane. Therefore, fundamental strings pull both probe brane and baryon vertex both until that the forces form two branes are balanced. The finial configurations for each case are drawn in Figure \ref{fig:D4D603}.

Up to now, we discuss $D$ brane configurations for single probe $D6$ brane. To discuss symmetry energy, we need at least two flavors. In this work, we consider two flavor (up and down) system and for simplicity we assume that masses of up and down quarks are the same. In $D$ brane point of view, we put two probe $D6$ brane on top of each other. 

The free energy of the system can be obtained by taking Legendre transformation of DBI action 
\be
{\cal F}_{D6} = \tilde{F} \frac{\partial S_{D6}}{\partial \tilde{F}} - S_{D6},
\ee
where $\tilde{F}$ is normalized field strength in (\ref{Sdbi}). We have two probe $D6$ branes and baryon vertex in nuclear matter system, we can calculate free energy of each probe brane by substituting the solution of equation of motion for each brane.

The total free energy of the system can be written as 
\be
{\cal F}_{\rm total}(Q) ={\cal F}_0 \,(Q) +{\cal F}_{D6}^{(1)}\,(Q_1) + {\cal F}_{D6}^{(2)}(Q_2),
\label{free}\ee
where $Q$ is number density of source and  ${\cal F}(Q)_0$ is zero for quark matter case and free energy of baryon vertex in nuclear matter case. Notice that this values only depends on total density and hence it does not contribute to calculation of symmetry energy.  We can define total charge  density and asymmetry parameter as 
\be
Q=Q_1 + Q_2,~~~~~\tilde{\a}=\frac{Q_1 - Q_2}{Q}.
\ee
If we fix the asymptotic value of two probe brane to be same, the total free energy has minimum at $\tilde{\a} =0$\cite{KSS2010}. Then we can  expand  total free energy in 
$\tilde{\a}$;
\be\label{ham03}
{\cal F}_{\rm total}(Q) = E_0 +E_1 \, \tilde{\a}  + E_2  \,\tilde{\a}^2 + \cdots.
\ee
The first term,  $E_0 = {\cal F}(Q)+2 {\cal F}_{D6}\left(\frac{Q}{2}\right)$,
can be identified with the free energy for symmetric matter. The second term in (\ref{ham03}) is zero because (\ref{free}) is symmetric in $Q_1$, $Q_2$.  
 The symmetry energy  is defined from the energy per nucleon and  given by
 \be\label{E2} 
   S_2 =\frac{E_2(Q)}{Q}  = \left(\frac{Q}{4}\right) \frac{\partial^2{\cal F}_{D6}^{(1)}(Q_1)}{\partial Q_1^2} \Bigg|_{Q_1 =Q/2} .
\ee
From (\ref{E2}), if we can write down the free energy for symmetric matter system in terms of density. We can calculate the symmetry energy by differentiating the total free energy with respect to density twice.

\subsection{Symmetry energy for nuclear matter system}
In this section, we discuss symmetry energy in nuclear matter system which corresponds to Figure \ref{fig:D4D603} (b).

\begin{figure}[!ht]
\begin{center}
\subfigure[]{\includegraphics[angle=0, width=0.47\textwidth]{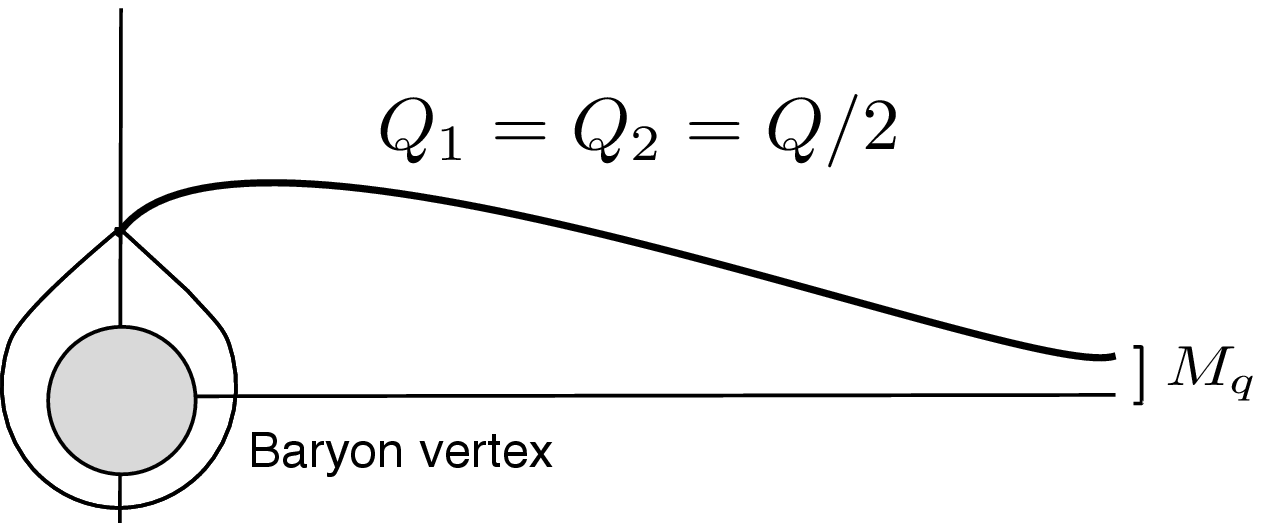}}
\subfigure[]{\includegraphics[angle=0, width=0.47\textwidth]{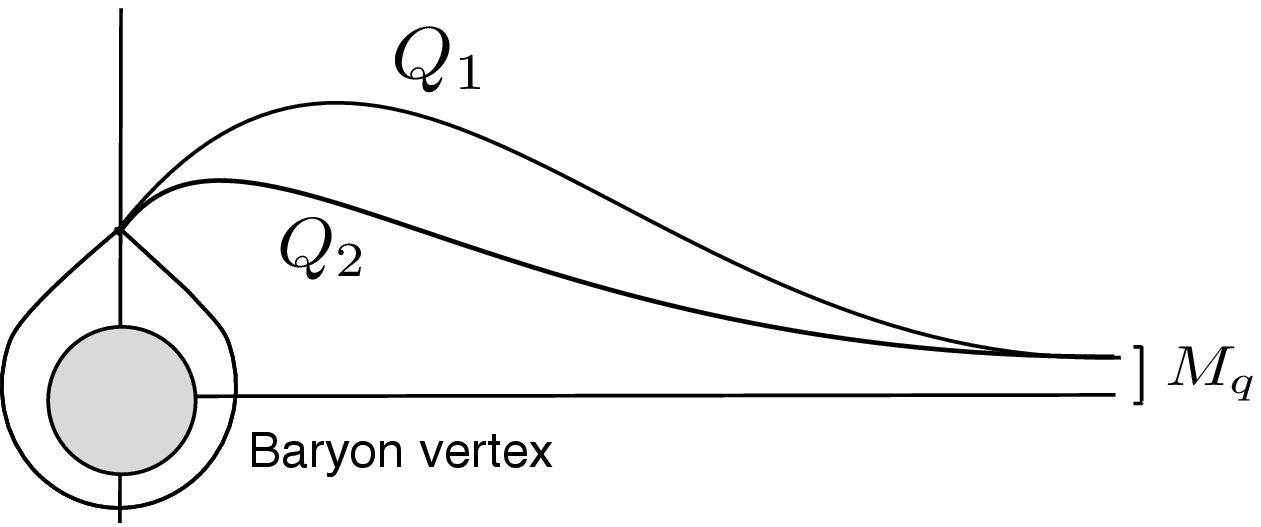}}
\caption{Schematic figure of D-brane configuration for (a) symmetric matter and (b) asymmetric matter. \label{fig:Sconf}}
\end{center}
\end{figure}

The symmetry energy can be understood as the energy costs when the system is deviated from symmetric states.  To achieve the deviation, we need to put  different number of charges(strings)  on each brane, which  gives 
different embedding for each probe brane. If the number of charges on each brane is same, two probe brane should be on top of each other. This configuration corresponds to the symmetric matter. The symmetry energy (\ref{E2}) can be understood the energy difference of the D-brane system between symmetric and asymmetric distribution of source on probe branes. The schematic figure is drawn in Figure. \ref{fig:Sconf}. 

Practically, what we need for calculation of symmetry energy is the free energy for symmetric matter system.  We first solve the equation of motion for probe brane numerically with given charge and quark mass together with  force balance condition.  Then substituting the solution  to (\ref{E2}), we can calculate the symmetry energy.

\begin{figure}[!ht]
\begin{center}
\subfigure[]{\includegraphics[angle=0, width=0.45\textwidth]{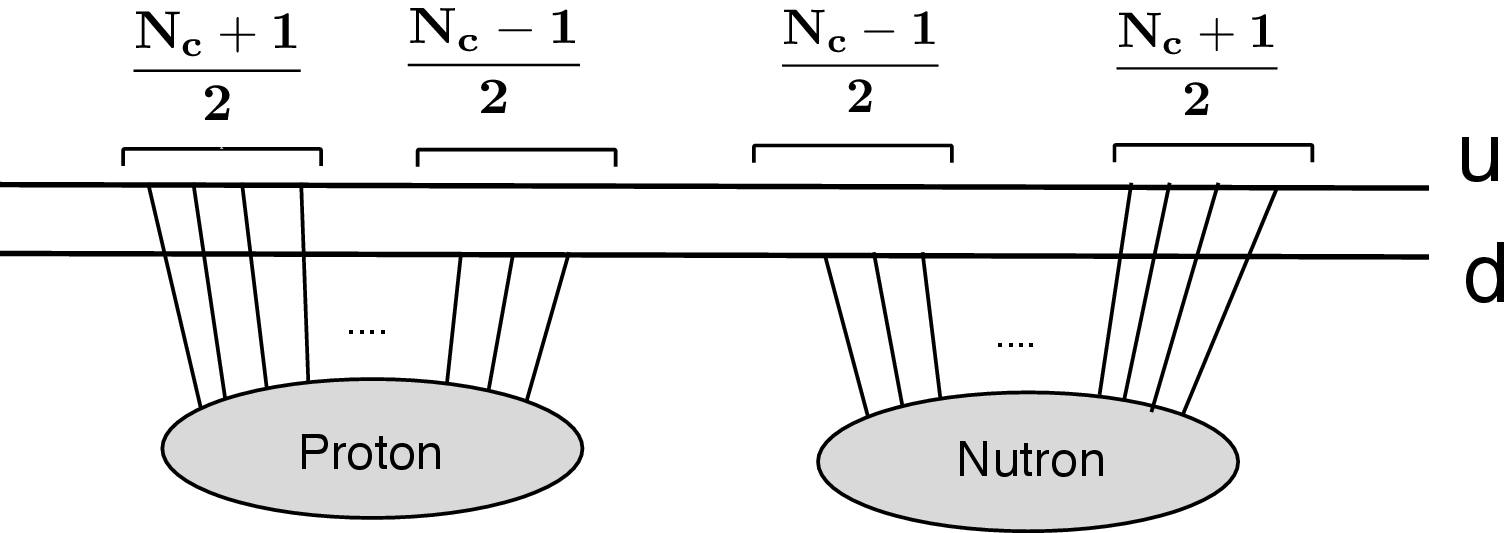}}
~~~~
\subfigure[]{\includegraphics[angle=0, width=0.45\textwidth]{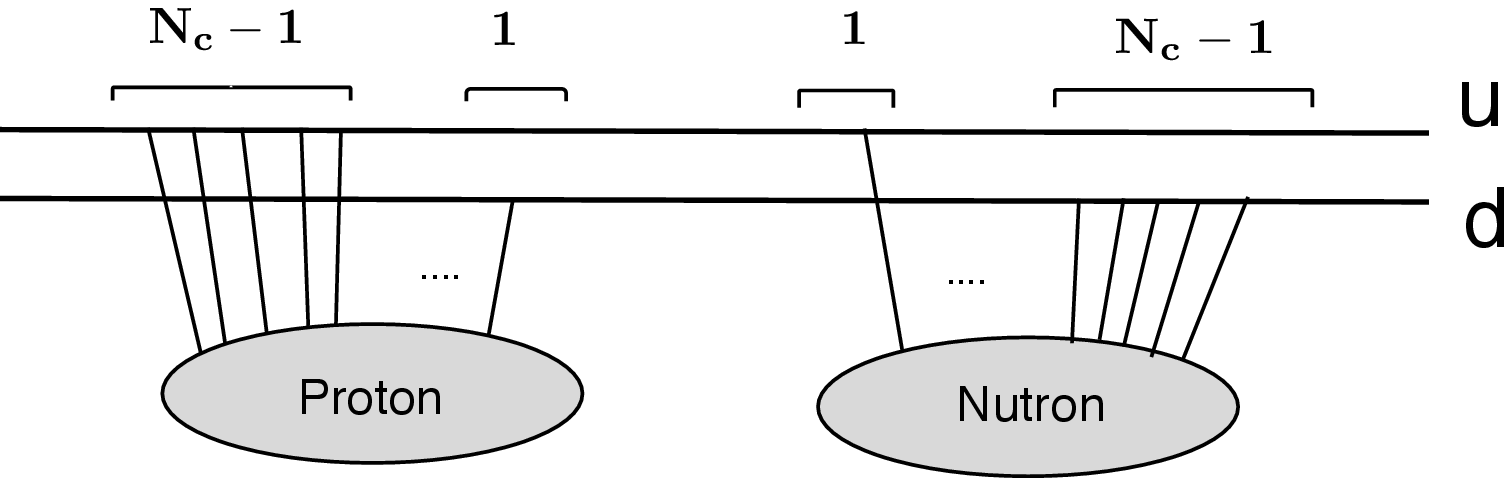}}
\caption{  proton and neutron in generic $N_c$. \label{fig:PN}}
\end{center}
\end{figure}

Before calculating symmetry energy, we need to define 
what is the proton and neutron for generic $N_C$ which is bigger than 3. In the case of $N_C =3$, proton consists of two up quarks and one down quark({\it uud}), and neutron is {\it udd}.  Among many possibilities for quark configurations of proton and neutron,
we consider  two possibilities in Figure \ref{fig:PN}. Notice that for discussing 
 nuclear matter, flavor number is 2 by definition. \footnote{ If we were interested in strange matter, we have to consider $N_f=3$.}
 
$\bullet$ case (a): 
In this case, the difference of quark number between proton and neutron is always $1$. To make this configuration be possible, we assume that $N_c$ is odd. From this configuration, we can set
\be
Q_1 - Q_2 = N_p - N_n, ~~~~Q_1 + Q_2 =Q =N_B \cdot N_C,
\ee
where $N_p$ is number of proton, $N_n$ is number of neutron and $N_B$ isn number of baryon i. e. $N_B =N_p +N_n$. Then, $\tilde{\a}$ can be written as
\be
\tilde{\a} =\frac{Q_1 - Q_2}{Q_1 + Q_2} \,=\,\frac{N_p -N_n}{N_C \cdot N_B}.
\ee
From this definition,  the second order term in (\ref{ham03}) becomes
\be
\tilde{\a}^2 E_2 = \left(\frac{N_p -N_n}{N_C \cdot N_B}\right)^2\cdot E_2  = \left(\frac{N_p -N_n}{N_B}\right)^2 \cdot \frac{E_2}{N_C^2}.
\ee
Then, the symmetry energy  per nucleon can be identified as
\be\label{E2a}
S_2 = \frac{E_2}{N_C^2 N_B}.
\ee
$E_2$ can be calculated from the free energy of probe $D6$ brane which has $N_C$ factor in the form. Therefore, 
there is $1/N_C$ factor in the symmetry energy (\ref{E2a}) which implies that the symmetry energy is suppressed by $N_C$. It is consistent with the definition of proton and neutron: there is only one quark difference between proton and neutron and therefore, for large $N_C$, it is not easy to distinguish these two particle and hence symmetry energy becomes zero for large $N_C$. 
\vskip0.1cm
$\bullet$ case (b): In this case, proton consist of $N_C -1$ up quarks and one down quark, and neutron has single up quark and $N_C -1$ down quark. The total difference and total number can be written in term of proton and neutron number as follows
\bea
Q_1 -Q_2 &=& (N_C -2)(N_p -N_n),\cr
Q_1 +Q_2 &=& N_C \cdot N_B.
\eea
Then,
\be
\tilde{\a} = \frac{Q_1 -Q_2}{Q_1+Q_2} = \frac{(N_C -2)\cdot(N_p -N_n)}{N_C N_B}.
\ee
The overall $N_C$ dependence of symmetry energy is
\be
S_2 \sim \frac{(N_C -2)^2}{N_C}.
\ee
Instead of suppression by $N_C$, the symmetry energy grows with $N_C$ factor for large $N_C$ limit. 

Considering other intermediate case in similar fashion, we can easily see that 
the free energy should be lowest in the case (a). Therefore we take the definition of 
proton defined in (a).  

\vskip0.2cm
 We can convert all parameters in terms of physical quantities such as 't Hooft coupling $\lambda$, Kaluza-Klein scale $M_{KK}$ and density. The nuclear density and quark mass can be written as
\bea
\varrho &=&\frac{Q}{N_c V_3} =\frac{2^{2/3} \,\Omega_2}{3^{4}\cdot (2\pi)^{4}} \lambda \,M_{KK}^3 \,\hat{Q},
\cr\cr
m_q &=&\frac{\lambda\, M_{KK}\, Y_{\infty}}{2^{2/3} \cdot 9\pi},
\eea
where $Y_{\infty}$ is asymptotic hight of probe $D6$ brane.
 By substituting these numerical solution into (\ref{E2}) we can get symmetry energy for each embeddings in terms of density and quark mass. From the meson mass calculation, we choose 
 \be 
 \lambda =18,~~ M_{KK} =1.04 \, GeV, ~~  N_C=3. 
 \ee
 
\begin{figure}[!ht]
\begin{center}
\includegraphics[angle=0, width=0.5\textwidth]{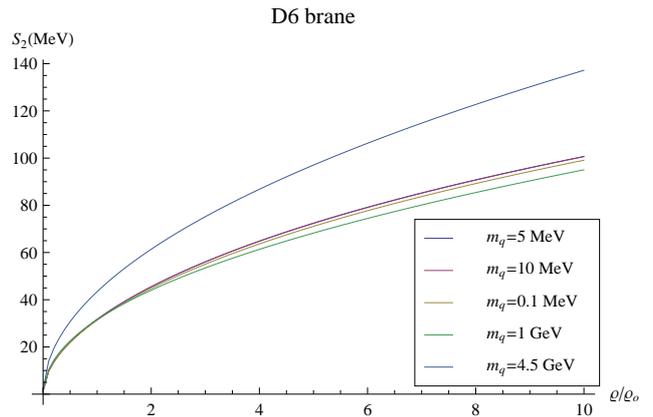}
\caption{Density dependence of symmetry energy for several quark mass for. \label{fig:S2_D6}}
\end{center}
\end{figure}

With this values, the numerical results of the symmetry energy are drawn in  Figure \ref{fig:S2_D6}.

From the Figure \ref{fig:S2_D6}, the symmetry energy with D6 probe brane seems to have square root behavior.   All the lines are well fitted to $S_2 =S_0 (\varrho/\varrho_0)^{1/2}$, with  $27 {\rm MeV} \le S_0 \le 36 {\rm MeV}$. For small quark mass($5 {\rm MeV}$), symmetry energy curve is fitted to $S_2 \sim 28 ({\rm MeV}) (\varrho/\varrho_0)^{1/2}$. As quark mass increase, the symmetry energy curve move downwards, in other words, it become softer up to quark mass is around 100 MeV. After then, the symmetry energy curve moves to upwards(stiffer) as quark mass increases.

For other configuration like $D3/D7$ or other probe brane, see \cite{Seo:2012sd}.

\subsection{Symmetry energy in quark matter system}
In this section we  consider symmetry energy in quark matter system which corresponds to Figure \ref{fig:D4D603} (a). This system has finite temperature and physical object is freely moving quark. Therefore, the boundary system is expected to quark gluon plasma. 

\begin{figure}[!ht]
\begin{center}
\subfigure[]{\includegraphics[angle=0, width=0.44\textwidth]{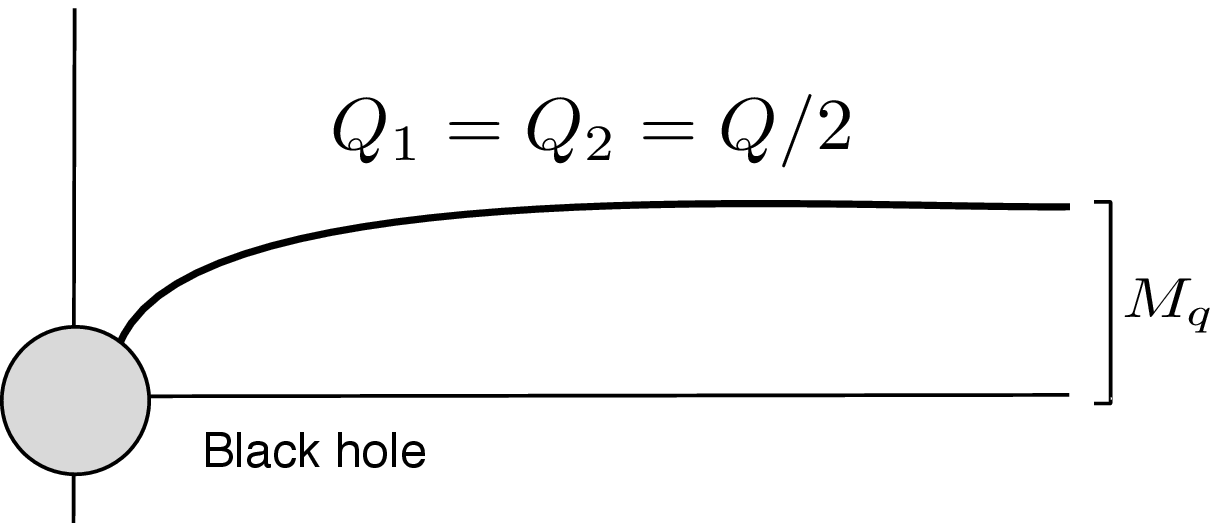}}
~~~~
\subfigure[]{\includegraphics[angle=0, width=0.44\textwidth]{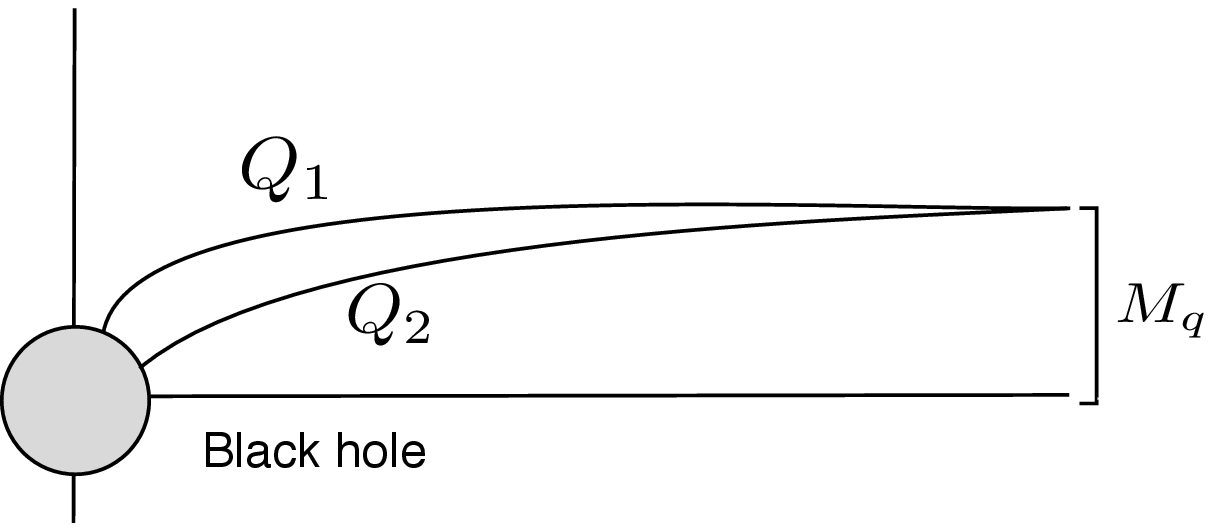}}
\caption{Schematic figure of D-brane configuration for (a) symmetric matter and (b) asymmetric matter. \label{fig:Sdeconf}}
\end{center}
\end{figure}

Similarly to the previous section, the symmetry energy can be understood  as the energy cost to separate the number of up and down quarks from symmetric matter. From the D-brane point of view, we need to consider two probe branes, 
where different number of strings are attached so that the embedding of  two branes are separated from each other.  The schematic figure is drawn in Figure \ref{fig:Sdeconf}. 

\begin{figure}[!ht]
\begin{center}
\includegraphics[angle=0, width=0.5\textwidth]{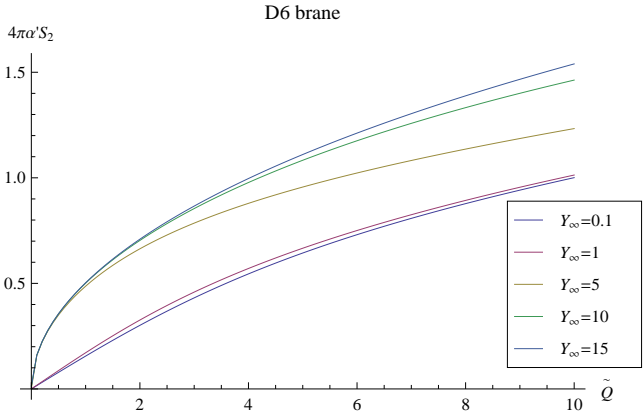}
\caption{Density dependence of symmetry energy for several $Y_{\infty}$. \label{fig:S2_D6bh}}
\end{center}
\end{figure}

By substituting the embedding solution into (\ref{E2}), we can get symmetry energy as a function of density and quark mass.
In this case $Y_{\infty}$, 
the asymptotic value of probe brane, which is related to  quark mass and temperature by 
\be
Y_{\infty} =\frac{2\pi l_s^2}{U_0} m_q =\frac{9 l_s}{8\pi R^3} \cdot \frac{m_q}{T^2}.
\ee
Therefore, if we fix quark mass, large value of $Y_{\infty}$ corresponds to low temperature and small $Y_{\infty}$ to high  temperature. 
The numerical results are drawn in Figure \ref{fig:S2_D6bh}.

In the figure, at low temperature(large value of $Y_{\infty}$),  the symmetry energy   $S_2 =0.5\, \tilde{Q}^{1/2}$, 
which is same as the case of nuclear matter system. 
But at high temperature(small value of $Y_{\infty}$), symmetry energy increase linearly in density. This linear behavior can be understood by the exact form of symmetry energy and it discussed  in \cite{Seo:2012sd} more preciesly.

\section{Scaling property and universality classes}
 In this section, we want to discuss scaling property of the symmetry energy from analytic calculation.
To do this, we consider the ideally simplified case: BPS metric background and flat embedding of probe branes. 
In this case,  background geometry becomes geometry of black $Dq$ brane;
\be\label{Dpmetric}
ds_{10}^2 = Z_p^{-1/2}(-dt^2 +d\vec{x}_p^2) +Z_p^{1/2} d\vec{x}_{\perp}^2,
\ee
with $e^{2\phi} = Z_p^{\frac{3-p}{2}}$, where $Z_p$ is harmonic function depends on dimensionality of background brane.
In generic background, the general form of the symmetry energy can be written in terms of the element of induce metric of probe brane as follows; 
\be\label{SSS}
S_2= 2\,\tau_q \int d\r \frac{\tilde{Q}\sqrt{G_{tt}G_{\r\r}}e^{-2\phi}\,G_{xx}^d G_{\O\O}^{n}}{\left(\tilde{Q}^2 +4 e^{-2\phi}\,G_{xx}^d G_{\O\O}^{n}\right)^{3/2}},
\ee
where $n \equiv q-d-1$ and $\rho$ is radial direction in $\vec{x}_{\perp}$. This result is derived in \cite{Seo:2012sd}.

From (\ref{Dpmetric}), the pre-factor of $G_{tt}$ is in verse of that of $G_{\rho\rho}$ and hence the square root term in numerator in (\ref{SSS}) is one. The other terms in (\ref{SSS}) becomes
\bea
e^{-2\phi} \, G_{xx}^d  \,G_{\Omega\Omega}^{q-d-1} 
&=&  \rho^{2n}\cdot Z_p^{\frac{p-3}{2}} \cdot Z_p^{-\frac{d}{2}}\cdot Z_p^{\frac{q-d-1}{2}}\cr\cr
&=& \rho^{2n} \cdot Z_p^{\frac{1}{2}(p+q-2d-4)}.
\eea
The value of $p+q-2d$ is precisely equal to the number of Neuman-Dirichlet (ND) direction  of $Dp/Dq$ system.  In the case of BPS $Dp/Dq$ system, half of supersymmetry is preserved and hence $p+q-2d =4$. Therefore, if we focus on the system which is supersymmetric configurations or  a smooth deformation of them,  the exponent becomes zero.  With this condition, we can get analytic result of the symmetry energy (\ref{SSS});
\bea\label{flatS2}
S_2 & =&2\,\tau_q \int d\r \frac{\tilde{Q} \r^{2 n}}{\left(\tilde{Q}^2 + 4\r^{2 n}\right)^{3/2}}
=c_n\,\tilde{Q}^{1/n},
\eea
where $c_n= 2\tau_q\frac{2^{-2-1/n}\Gamma\left(\frac{1}{2n}\right)\Gamma\left(\frac{n-1}{2n}\right)}{n^2 \sqrt{\pi}}$ and $n=q-d-1$ so that the density dependence of symmetry energy is
$ S_2 \sim Q^{\frac{1}{q-d-1}}$.  This results are summarized in Table 1 and Table 2.
We also check that  this result is consistent with the numerical calculation for small density region.

\begin{table}
\caption{$D4$ brane background}\label{S2D4}
\begin{tabular}{ |c||  c | c | c || c |c|}
\hline
   &   $q$ & $d$ & $q-d-1$ & $S_2$ &$2\nu =n/d$ \\
\hline
$D4$&  & & & &   \\
\hline \hline 
$D2$& 2&1&0&${\cal O}(1)$ &-\\
\hline
$D4$& 4&2&1&$Q$& $1/2$\\
\hline
$D6$& 6&3&2&$Q^{1/2}$ &$2/3$ \\
\hline
\end{tabular}
\end{table}

\begin{table}
\caption{$D3$ brane background}\label{S2D3}
\begin{tabular}{ |c||  c | c | c || c |c|}
\hline
   &   $q$ & $d$ & $q-d-1$ & $S_2$ &$2\nu =n/d$ \\
\hline
$D4$&  & & & &   \\
\hline \hline 
$D2$& 3&1&1& $Q$  &1\\
\hline
$D4$& 5&2&2&$Q^{1/2}$& 1\\
\hline
$D6$& 7&3&3&$Q^{1/3}$ &1 \\
\hline
\end{tabular}
\end{table}

Notice that both  the background geometry and embedding used here  are not exact dual of the real QCD: 
real background is a deformation of such BPS solution 
and the embedding is non-trivial deformation from such a flat embedding. 
Nevertheless, the scaling exponent  is expected to be same as the actual configuration. 
The point  is that  smooth change  of the metric or the 
embedding shape does not seem to  change the scaling behavior of the symmetry energy. 
The exponent of symmetry energy  depends only  on the dimensionality of 
probe brane and dimension of non-compact directions. 
{\it Therefore the scaling exponents depend only on the universality classes. }

\section{Discussion}
In this review, we demonstrated how to calculate the symmetry energy for  both nuclear matter using the gauge/gravity duality. 
The symmetry energy has a power like density dependence with characteristic exponent 
which is invariant under the smooth deformation of the metric as well as smooth deformation of the 
embedding.  
Therefore it is a index for the universality class. 
The physical interpretation of the scaling exponent is still open question. A trial interpretation as the 
 Non-fermi liquid nature of the nuclear matter is given in the original paper
 \cite{Seo:2012sd}.

\section*{Acknowledgments}
This work was supported by the  NRF grant funded by the Korea government(MEST) through the  Mid-career Researcher Program  with grant NRF-2013R1A2A2A05004846. The work was partially supported by the WCU project of Korean Min- istry of Education, Science and Technology (R33-2008-000-10087-0). 
 The work of YS was partly supported by 
Basic Science Research Program through the National Research Foundation of Korea(NRF) funded by 
the Ministry of Education(NRF-2012R1A1A2040881).

\end{document}